\documentclass[multphys,vecphys]{svmult}

\usepackage{makeidx}         
\usepackage{graphicx}        
\usepackage{multicol}        
\usepackage[bottom]{footmisc}
\usepackage{bm}
\usepackage{epsf}

\makeindex             

\newcommand{\lambdabar}{{\hbox{$\lambda$\kern-1.ex\raise+0.45ex\hbox{--}}}}

\begin{document}

\title*{Relic neutrino clustering  and implications for their detection\protect\footnote{Talk given by YYYW at DARK2004,
College Station TX, USA.
Based on \cite{paper}.}}
\titlerunning{Relic neutrino clustering  and implications for their detection}
\author{Andreas Ringwald \and Yvonne~Y.~Y.~Wong}
\institute{Deutsches Elektronen-Synchrotron DESY, Hamburg, Germany
\texttt{andreas.ringwald@desy.de}, \texttt{yvonne.wong@desy.de}}
%
%

\maketitle

\vspace*{-2.2in}
\rightline{\large\tt DESY 04-242}
\vskip 2.0in

\begin{abstract}
We study the gravitational clustering of
big bang relic neutrinos onto existing cold dark matter and baryonic
structures within the flat $\Lambda$CDM model.
We then discuss the implications of clustering for
scattering-based relic neutrino detection methods,
ranging from flux detection via Cavendish-type torsion balances, to target
detection using accelerator
beams and cosmic rays.

\end{abstract}

\section{Introduction}
\label{intro}

The standard big bang theory predicts the existence of
$10^{87}$ neutrinos per
flavour in the visible universe.
This is an enormous abundance unrivalled
by any other known form of matter, falling second only to the cosmic microwave
background (CMB) photon.  Yet, unlike the CMB which boasts its first
 detection in the 1960s and which has since been observed
and its properties
measured to high accuracy in a series of  experiments,
the relic neutrino continues to be elusive in the laboratory.
The chief reason for this is of course the feebleness of the weak
interaction.  The smallness of the neutrino mass also makes
momentum-transfer-based detection methods highly impractical.
At present, the only evidence for the relic neutrino comes from
inferences
from other cosmological measurements, such as big bang nucleosynthesis,
CMB and large scale structure (LSS) data (e.g.,
 \cite{bib:hannestadreview}).
Nevertheless, it is difficult to accept that these neutrinos will never be
detected in a more direct way.

In order to design feasible direct, scattering-based detection methods,
a precise knowledge
of the relic neutrino phase space distribution  is indispensable.
In this connection, it is important note that an oscillation interpretation
of the atmospheric
and solar neutrino  data (e.g.,
 \cite{bib:atmospheric})
implies that at least two
of the neutrino mass eigenstates are nonrelativistic today. These neutrinos
are subject to gravitational clustering on existing cold dark
matter (CDM) and baryonic structures,
possibly causing the local neutrino number
density to depart from the standard value of
$\bar{n}_{\nu} = \bar{n}_{\bar{\nu}} \simeq 56 \ {\rm cm}^{-3}$,
and the momentum
distribution to deviate from the relativistic Fermi--Dirac function.

In this talk, we describe a method that allows us to study
the gravitational clustering of relic neutrinos onto CDM/baryonic structures.
We calculate the present day neutrino overdensities in general
CDM halos and in the Milky Way, and then discuss their implications for
scattering-based relic neutrino detection methods---
from flux detection via Cavendish-type torsion balances, to target
detection using accelerator
beams and cosmic rays.

\section{\label{vlasov} Vlasov equation}

The standard procedure for any clustering investigation involving  gravity
 only is to solve the Vlasov, or collisionless Boltzmann,
equation
(e.g., \cite{bib:bertschinger,bib:klypin}),
\begin{equation}
\label{eq:vlasov} \frac{D f_i}{D \tau} \equiv \frac{\partial
f_i}{\partial \tau} + \dot{\bm x} \cdot \frac{\partial
f_i}{\partial {\bm x}} + \dot{\bm p} \cdot \frac{\partial
f_i}{\partial {\bm p}} = 0.
\end{equation}
The single-particle phase density $f_i({\bm x}, {\bm p}, \tau)$ is
defined so that $d N_i=f_i \ d^3x \  d^3p$ is the number of $i$ type
particles (e.g., CDM, neutrinos) in an infinitesimal phase space volume element.  The
variables
${\bm x} = {\bm r}/a(t)$,
${\bm p} = a m_i \dot{\bm x}$, and $d \tau = dt /a(t)$
are the comoving distance, its associated conjugate momentum, and
the conformal time  respectively, with  $a$ as the scale factor and
$m_i$ the mass of the $i$th particle species.
All temporal and spatial
derivatives are taken with respect to comoving coordinates,
i.e.,
$\dot{}\equiv \partial/\partial \tau$, $\nabla \equiv
\partial/\partial {\bm x}$.\footnote{Unless otherwise indicated,
comoving spatial and temporal quantities are used throughout the present work.
Masses and densities, however, are always physical.}

In the nonrelativistic, Newtonian limit, equation (\ref{eq:vlasov}) is
equivalent to
\begin{equation}
\label{eq:vlasov2} \frac{\partial f_i}{\partial \tau} + \frac{\bm
p}{a m_i} \cdot \frac{\partial f_i}{\partial {\bm x}} - a m_i
\nabla \phi \cdot \frac{\partial f_i}{\partial {\bm p}} = 0,
\end{equation}
with the Poisson equation
\begin{equation}
\nabla^2 \phi = 4 \pi G a^2 \sum_i \overline\rho_i(\tau)
\delta_i ({\bm x}, \tau), \label{eq:poisson}
\end{equation}
\begin{equation}
\delta_i(\bm{x},\tau) \equiv   \frac{\rho_i({\bm x}, \tau)}{
\overline\rho_i(\tau)} -1,  \qquad \rho_i (\bm{x},\tau) =
\frac{m_i}{a^3} \int d^3p \ f_i (\bm{x},\bm{p}, \tau), \label{eq:fluctuations}
\end{equation}
relating the peculiar gravitational potential $\phi (\bm{x},\tau)$
to the density fluctuations $\delta_i (\bm{x},\tau)$ with respect
to the physical mean $\bar{\rho}_i(\tau)$.

The Vlasov equation expresses conservation of phase space density $f_i$
along each characteristic $\{{\bm x}(\tau),{\bm p}(\tau)\}$ given by
\begin{equation}
\label{eq:characteristic}
\frac{d \bm{x}}{d \tau} = \frac{\bm{p}}{a m_i}, \qquad \frac{d \bm{p}}{d \tau}
= - a m_i \nabla \phi.
\end{equation}
The complete set of characteristics coming through every point in phase 
space is thus exactly equivalent to equation (\ref{eq:vlasov}).
It is generally
not possible to follow the whole set of characteristics, but the evolution
of the system can still be traced, to some extent, if we follow a
sufficiently large but still manageable sample selected from the
initial phase space distribution.  This forms the basis of particle-based
solution methods.

\section{\label{profiles} Solution method and halo density profiles}

A ``first principles'' approach to neutrino clustering
requires the simultaneous solution of the Vlasov equation (\ref{eq:vlasov})
for both CDM and neutrinos.  This is usually done by means of
multi-component $N$-body simulations.
In our treatment, however, we make two simplifying approximations:
\begin{enumerate}
\item
We assume only the CDM component $\rho_m$ contributes to $\phi$ in the
Poisson equation (\ref{eq:poisson}), and $\rho_m$ to be
completely specified by halo density profiles from high resolution
$\Lambda$CDM simulations \cite{bib:singh&ma}.  The neutrino component
is treated as a small perturbation whose clustering depends on the
CDM halo profile, but is too small to affect it in return.
This assumption is well justified, since, on cosmological scales,
LSS data require
$\rho_{\nu}/\rho_m = \Omega_{\nu}/\Omega_{m} < 0.2$ \cite{bib:hannestadreview}.
On cluster/galactic scales, neutrino free-streaming ensures
that $\rho_{\nu}/\rho_m$ always remains smaller than $\Omega_{\nu}/\Omega_{m}$
\cite{bib:kofman}.

\item Given that assumption 1.\ holds, it follows that
not only will the CDM halo be gravitationally
blind to the neutrinos, the neutrinos themselves will also have
negligible gravitational interaction with each other.
\end{enumerate}
These approximations together allow us to track the neutrinos one
at a time in $N$ independent simulations, instead of following $N$ particles
simultaneously in one single run.  We shall call this ``$N$-one-body simulation'' \cite{paper}.

For the halo density profiles, we use the ``universal
profile'' advocated by Navarro, Frenk and White (hereafter, NFW)
\cite{Navarro:1995iw,bib:nfw},
\begin{equation}
\label{eq:nfw}
\rho_{\rm halo}(r) = \frac{\rho_s}{(r/r_s) (1 + r/r_s)^2}.
\end{equation}
The parameters $r_s$ and $\rho_s$
are determined by the halo's virial mass $M_{\rm vir}$ and a
dimensionless concentration
parameter
$c \equiv r_{\rm vir}/r_s$,
where $r_{\rm vir}$ is the virial radius, within which lies $M_{\rm vir}$
of matter with an average density equal to $\delta_{\rm TH}$ times
the mean matter density  $\bar{\rho}_m$ at that redshift, i.e.,
\begin{equation}
\label{eq:mvir}
M_{\rm vir}  \equiv   \frac{4 \pi}{3} \delta_{\rm TH}
\bar{\rho}_m a^3 r_{\rm vir}^3
= \frac{4 \pi}{3} \delta_{\rm TH}
\bar{\rho}_{m,0} r_{\rm vir}^3,
\end{equation}
where $\bar{\rho}_{m,0}$ is the present day mean matter density.
The factor $\delta_{\rm TH}$ is the overdensity
predicted by the dissipationless spherical top-hat collapse model,
\begin{equation}
\delta_{\rm TH} \simeq \frac{18 \pi^2 + 82 y - 39 y^2}{\Omega_m(z)}, \quad
y = \Omega_m (z) -1,
\end{equation}
with  $\Omega_m(z) = \Omega_{m,0}/(\Omega_{m,0} + \Omega_{\Lambda,0} a^3)$
\cite{bib:bryan&norman}.

Furthermore, halo concentration correlates with its mass.
At $z=0$, the trend
\begin{equation}
\label{eq:cmvir}
c(z=0) \simeq 9 \left(\frac{M_{\rm vir}}{1.5 \times 10^{13} h^{-1} M_{\odot}}
\right)^{-0.13}
\end{equation}
was found in \cite{bib:bullock2001}.   In addition, for a fixed virial mass, the median
concentration parameter exhibits a redshift dependence of
$c(z) \simeq c(z=0)/(1+z)$
between $z=0$ and $z=4$.

\section{Clustering in NFW halos}

Using the NFW halo profile (\ref{eq:nfw}) as an input,
we find solutions to the Vlasov equation in the limit
$\rho_{\nu} \ll \rho_m$.
The CDM distribution is modelled as follows: We assume a uniform distribution of CDM throughout space, with
a spherical NFW halo sitting at the origin. For the neutrinos, we take their initial distribution
to be the homogeneous and isotropic Fermi--Dirac
distribution with no chemical potential.
The initial redshift is taken to be $z=3$, since, at higher redshifts, a sub-eV neutrino has
too much thermal velocity to cluster efficiently. The cosmological parameters used are
$\{\Omega_m,\Omega_{\Lambda},h\}= \{0.3,0.7,0.7\}$.

\begin{figure}[t]
\centering
\includegraphics[width=11.5cm]{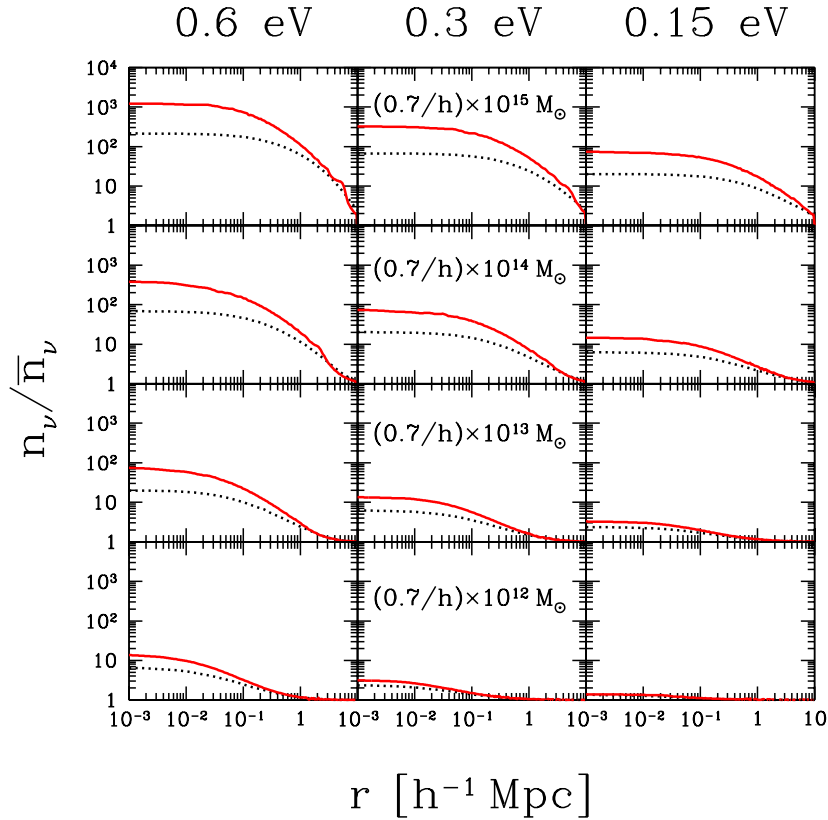}
%
%
\caption{\label{fig:overdensities}Relic neutrino number density per
flavour, $n_{\nu}=n_{\bar{\nu}}$, normalised
to $\bar{n}_{\nu}=\bar{n}_{\bar{\nu}}  \simeq 56 \ {\rm cm}^{-3}$, for the
indicated
neutrino and halo virial masses.  Results from $N$-one-body simulations are
denoted by red (solid) lines.  Dotted lines correspond to overdensities
calculated with the linear approximation.}
\end{figure}

\begin{figure}[t]
\centering
\includegraphics[width=11.5cm]{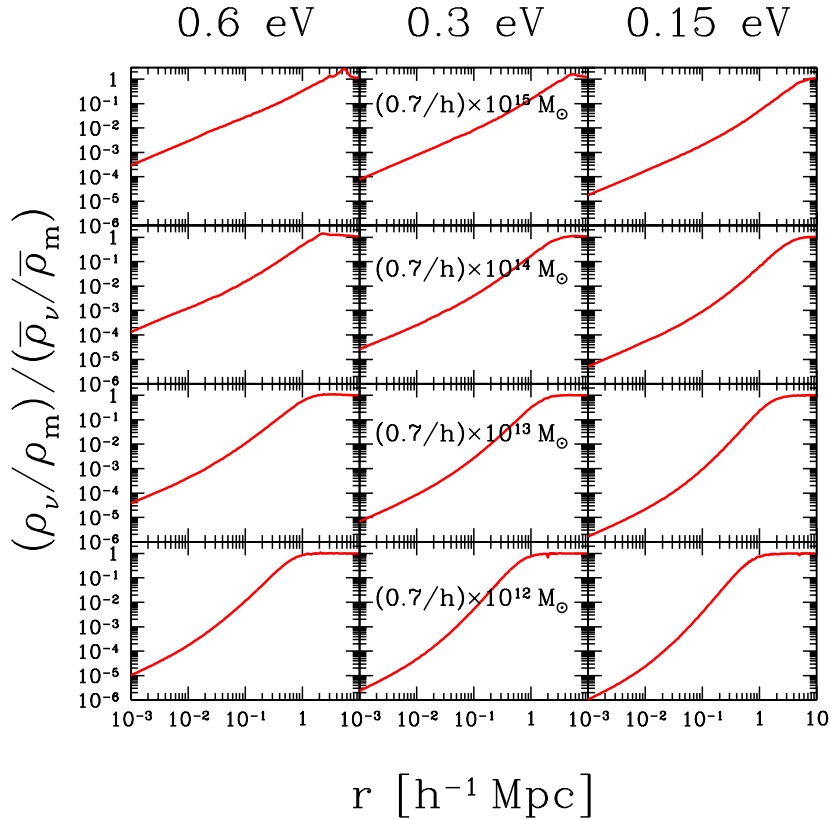}
%
%
\caption{\label{fig:nuvscdm}Mass density ratio $\rho_{\nu}/\rho_{m}$
normalised to the background mean $\bar{\rho}_{\nu}/\bar{\rho}_m$
obtained from $N$-one-body simulations for the indicated
neutrino and halo masses.}
\end{figure}

We solve the Vlasov equation using $N$-one-body simulations,
as well as a semi-analytical linear method.\footnote{The linear approximation \cite{bib:gilbert} consists of
replacing $\partial f/\partial {\bm p}$ with $\partial f_0/\partial {\bm p}$ in
(\ref{eq:vlasov}), where $f_0$ is the unperturbed Fermi--Dirac function.}
The essential features  of the results (Figures
\ref{fig:overdensities} and \ref{fig:nuvscdm})
can be understood
in terms of neutrino free-streaming, which causes
$n_{\nu}/\bar{n}_{\nu}$  to flatten out at small radii, and
the mass density ratio $\rho_{\nu}/\rho_{m}$ to drop substantially
below the background mean.
Both $n_{\nu}/\bar{n}_{\nu}$ and $\rho_{\nu}/\rho_m$
approach their respective
cosmic mean of $1$ and $\bar{\rho}_{\nu}/\bar{\rho}_m$
at large radii.

Furthermore, we find that the linear method systematically
underestimates the neutrino overdensities over the whole range of
halo and neutrino masses considered here.
Reconciliation with $N$-one-body simulations can only be achieved if
we impose a smoothing scale of $> 1 \ {\rm Mpc}$, or if
$n_{\nu}/\bar{n}_{\nu} < 3 \div 4$.
This finding is consistent with the standard lore that perturbative methods fail once
the perturbations exceed unity and nonlinear effects set in.

\begin{figure}[t]
\centering
\includegraphics[width=11.5cm]{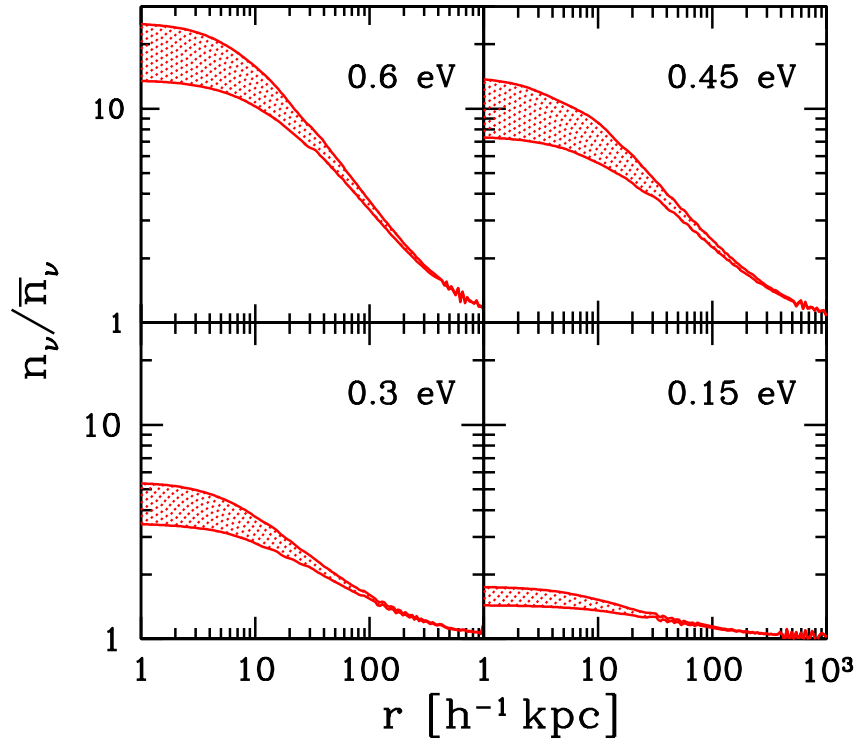}
%
%
\caption{\label{fig:milkyway}Relic neutrino number density per
flavour, $n_{\nu}=n_{\bar{\nu}}$,
 in the Milky Way for various neutrino masses.
All curves are normalised to
$\bar{n}_{\nu} = \bar{n}_{\bar{\nu}} \simeq 56 \ {\rm cm}^{-3}$.  The
top curve in each plot corresponds
to the MWnow run, and the bottom to the NFWhalo run.
The enclosed region represents a possible range of
overdensities at $z=0$.}
\end{figure}

\section{Clustering in the Milky Way}

In order to calculate the neutrino overdensity in the Milky Way and, especially,
their phase space distribution at Earth
($r_{\oplus}
\sim 8 \ {\rm kpc}$ from the Galactic Centre), we need, in principle, to know
the complete assembly history of the Milky Way.   Theory
suggests that the galactic bulge and disk grew out of an NFW halo via baryonic
compression \cite{bib:white&rees,bib:mmw}.  Our strategy, then, is to conduct two series of simulations,
one for the present day Milky Way mass distribution (MWnow) \cite{bib:dehnen,bib:klypin2002}
which we assume to be static, and one for the NFW halo (NFWhalo)
that would have been there, had baryon compression not taken place.
The real neutrino overdensity should then lie somewhere
between these two extremes.  Figure \ref{fig:milkyway} shows the
possible ranges of
overdensities at $z=0$.

\begin{figure}[t]
\centering
\includegraphics[width=11.5cm]{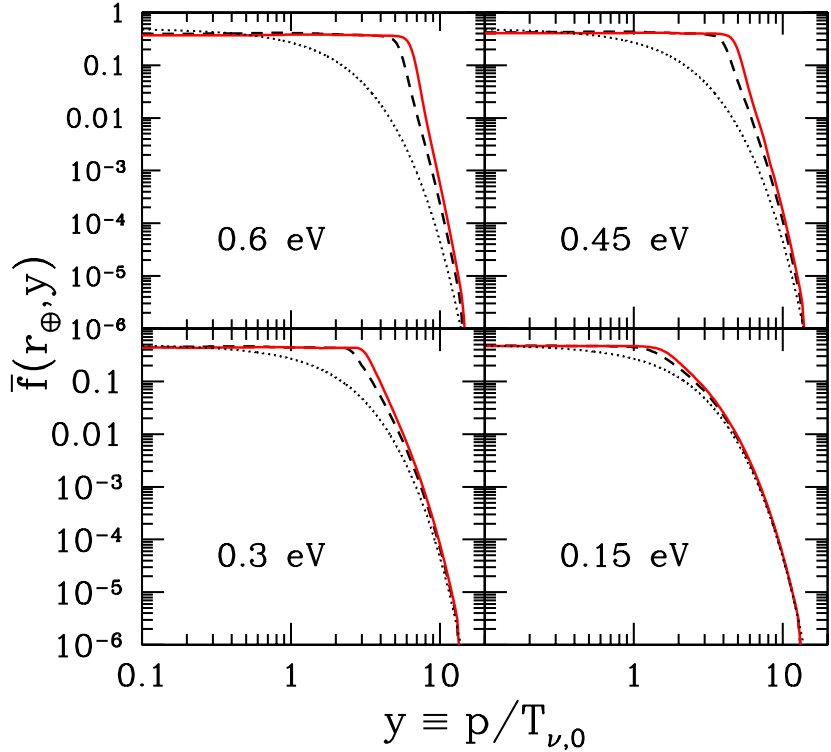}
%
%
\caption{\label{fig:momentum}Momentum distribution of relic neutrinos
at $r_{\oplus}$ for various neutrino masses.
The red (solid) line denotes  the MWnow run,
while the dashed line represents the NFWhalo run.  The relativistic
Fermi--Dirac function is indicated by the dotted line.  The
escape velocity $v_{\rm esc} = \sqrt{2 |\phi(r_{\oplus})|}$ is
$490 \ {\rm km \ s}^{-1}$ and $450 \ {\rm km \ s}^{-1}$ for MWnow and
NFWhalo respectively, corresponding to ``escape momenta''
$y_{\rm esc} \equiv m_{\nu} v_{\rm esc}/T_{\nu,0}$
of $(5.9,4.4,3.0,1.5)$ and $(5.4,4.1,2.7,1.4)$ for
$m_{\nu} = (0.6,0.45,0.3,0.15) \ {\rm eV}$.}
\end{figure}

In all cases, the final momentum distribution at $r_{\oplus}$ is almost
isotropic, with a zero mean radial velocity $\langle v_r \rangle$,
and second velocity
moments that satisfy approximately the relation $2 \langle v_r^2 \rangle =
\langle v_T^2 \rangle$.  Hence, we plot the coarse-grained
phase space densities $\bar{f}(r_{\oplus},p)$ only as
functions of the absolute velocity.

The coarse-grained spectra in Figure \ref{fig:momentum}
show varying degrees of deviation from the relativistic Fermi--Dirac function,
but share a common feature that $\bar{f} \sim 1/2$  up to the
momentum state corresponding to the escape velocity from the
Milky Way at $r_{\oplus}$.  This agrees with the requirement
that the final coarse-grained density
must not exceed the maximal initial fine-grained
distribution,  $\bar{f} \leq \max(f_0)$ \cite{bib:lyndenbell,bib:tremaine&gunn,bib:shu1978,bib:shu1987,bib:kull}.
For neutrinos, $\max(f_0) = 1/2$ at $p=0$.  Thus, our $\bar{f}$ not only satisfies but completely
saturates the bound up to $p_{\rm esc}$, forming
a semi-degenerate state that can only be made denser by filling
in states above $p_{\rm esc}$.\footnote{This degeneracy should not be confused with that arising from
the Pauli exclusion
principle.}

\section{Relic neutrino detection}

\subsection{Flux detection}

The relic neutrinos' low average momentum $\langle p\rangle =\langle y\rangle\, T_{\nu ,0}$
 corresponds to a de Broglie wavelength
of macroscopic dimension,
$\lambdabar = 1/\langle p\rangle = 0.12$~cm$/\langle y\rangle$.
Therefore, one may envisage scattering processes in which many target atoms
act
coherently~\cite{Shvartsman:sn,Smith:jj} over a macroscopic volume
$\lambdabar^3$, so that the elastic scattering rate   is
proportional to the square of the number of target atoms in $\lambdabar^3$.
Compared to the case where the neutrinos are
elastically scattered coherently only on
the individual target nuclei,
the new rate is enhanced by a factor of
\begin{equation}
\label{eq:enhancement}
\frac{N_A}{A}\,\rho_{\rm t}\,\lambdabar^3
\simeq 6\times 10^{18}\,\left( \frac{100}{A}\right)
\left( \frac{\rho_{\rm t}}{{\rm g/cm^3}}\right)
\left( \frac{\lambdabar}{0.1\ {\rm cm}}\right)^3
\,,
\end{equation}
where $N_A$ is the Avogadro constant, $A$ is the atomic mass, and
$\rho_{\rm t}$ is the
mass density of the target
material.
\footnote{In the case of coherent scattering,
it is possible, in principle, to measure also the scattering amplitude
itself~\cite{Stodolsky:1974aq,Cabibbo:bb,Langacker:ih}, which
is linear in $G_F$.
However, a large lepton asymmetry is required for
a non-negligible  effect.}

Exploiting this effect, a practical
detection scheme for the local relic neutrino flux
is based on the fact that a test body of density $\rho_{\rm t}$
at Earth experiences a neutrino wind force through random
scattering events, leading
to an acceleration  given, for Dirac neutrinos, by \cite{Shvartsman:sn,Smith:jj,Duda:2001hd,Ferreras:wf}
\begin{eqnarray}
\label{eq:accel}
a_{\rm t} &\simeq & \sum_{\nu,\bar\nu}\
\underbrace{n_{\nu}\,v_{\rm rel}}_{\rm flux}\
\frac{4\pi}{3}\, N_A^2\, \rho_{\rm t}\,  r_{\rm t}^3
\
\sigma_{\nu N}\,
\underbrace{2\,m_\nu\,v_{\rm rel}}_{\rm mom.\, transfer}
\nonumber \\
&\simeq &
2\times 10^{-28}\
\left( \frac{n_\nu}{\bar n_\nu}\right)
\left( \frac{10^{-3}\,c}{v_{\rm rel}}\right)
\left( \frac{\rho_{\rm t}}{{\rm g/cm^3}}\right)
\left( \frac{r_{\rm t}}{\hbar/(m_\nu v_{\rm rel})}\right)^3
 {\rm cm}\ {\rm s}^{-2},
\end{eqnarray}
with $r_t < \lambdabar$,   $\sigma_{\nu N}\simeq G_F^2\, m_\nu^2/\pi$ is the elastic
neutrino--nucleon cross section,
$v_{\rm rel}=\langle |\bm{v} - \bm{v}_\oplus|\rangle$  the mean
neutrino velocity
in the detector's rest frame, and $v_\oplus\simeq 2.3 \times 10^2\ {\rm km \ s}^{-1}
\simeq 7.7\times 10^{-4}\,c$
the Earth's velocity through the Milky Way.  For $n_{\nu}/\bar{n}_{\nu} \sim 20$,
equation (\ref{eq:accel}) gives $a_t \sim 10^{-26} \ {\rm cm} \ {\rm s}^{-2}$.
For Majorana neutrinos, $a_t$ is further suppressed
by a factor of
$(v_{\rm rel}/c)^{2}\simeq 10^{-6}$
for an unpolarised target, and $v_{\rm rel}/c\simeq 10^{-3}
$ for a polarised one.

To digest these estimates, we note that the smallest
measurable acceleration at present
is $> 10^{-13} \ {\rm cm \ s}^{-2}$, using
conventional Cavendish-type torsion balances.
Possible improvements with currently available
technology to a sensitivity of
$> 10^{-23} \ {\rm cm \ s}^{-2}$ have been
proposed \cite{Hagmann:1998nz,Hagmann:1999kf}. However, this
is still off the prediction~(\ref{eq:accel}) by
three orders of magnitude.
Therefore, we conclude that an observation of this effect will not be
possible in the next decade, but can still be envisaged in the foreseeable
future
(thirty to forty years according to
\cite{Smith:sy}, exploiting advances
in nanotechnology), if
our known light neutrinos are Dirac particles. Should they turn out,
in the meantime, to be
Majorana particles, flux detection via mechanical forces will be
a real challenge.

Lastly, the background contribution to the acceleration~(\ref{eq:accel})
from  the solar $pp$ neutrinos [${\rm flux} \sim 10^{11} \ {\rm cm}^{-2}
{\rm s}^{-1}$, $\langle E_{\nu} \rangle \sim 0.3 \ {\rm MeV}$ (e.g., \cite{bib:bahcall})],
$a_t^{\nu\,{\rm sun}}\simeq 10^{-27}\ {\rm cm \ s}^{-2}$~\cite{Duda:2001hd},
may be rejected by directionality.
The background from weakly interacting massive particles (WIMPs $\chi$, with mass $m_{\chi}$)~\cite{Duda:2001hd},
\begin{eqnarray}
\label{eq:accel_WIMP}
a_{\rm t}^{\rm WIMP} &\simeq &
\underbrace{n_{\chi}\,v_{\rm rel}}_{\rm flux}\
N_A\,A\
\sigma_{\chi N}\,
\underbrace{2\,m_\chi\,v_{\rm rel}}_{\rm mom.\, transfer}
\\ \nonumber
&\simeq &
6\times \! 10^{-29}
\left( \frac{\rho_\chi}{0.3\ {\rm GeV/cm^3}}\! \right) \!
\left( \frac{v_{\rm rel}}{10^{-3}\,c}\right)^2 \!
\left( \frac{A}{100}\right) \!
\left( \frac{\sigma_{\chi N}}{10^{-45}\ {\rm cm^2}}\! \right)
 {\rm cm}\ {\rm s}^{-2},
\end{eqnarray}
should
they be the main constituent of galactic dark matter with mass density
$\rho_\chi\equiv n_\chi m_\chi\simeq 0.3\ {\rm GeV \ cm}^{-3}$ at $r_\oplus$,
can be neglected as soon as the WIMP--nucleon cross section $\sigma_{\chi N}$
is smaller than $\sim 3\times 10^{-45}$~cm$^2$.
This should be well established by the time relic neutrino direct detection
becomes a reality.

\subsection{Target detection}

Detection methods based on the scattering of extremely energetic particles
(accelerator beams or cosmic rays) off the relic neutrinos as
a target take advantage of the fact that,
for centre-of-mass (c.m.) energies,
\begin{equation}
\sqrt{s}=  \sqrt{2\,m_\nu\,E_{\rm beam}}\simeq 4.5 \
\left( \frac{m_\nu}{\rm eV}\right)^{1/2}\,
\left( \frac{E_{\rm beam}}{\rm 10\ TeV} \right)^{1/2} \ {\rm MeV},
\end{equation}
just  below the $W$- and $Z$-resonances,
the weak interaction cross sections grow rapidly with the
beam energy $E_{\rm beam}$.

\subparagraph{\label{accelerator} At accelerators}

Target detection using accelerator beams does not seem viable.
For a hypothetical beam energy of $10^7 \ {\rm TeV}$ and
an accelerator ring of ultimate circumference $L\simeq 4\times 10^4 \ {\rm km}$ around the Earth,
the interaction rate is roughly one event per year.  See \cite{paper} for details.

\subparagraph{\label{uhecr} With cosmic rays}

It was pointed out  by
Weiler~\cite{Weiler:1982qy,Weiler:1983xx} (for earlier suggestions, see
\cite{Bernstein:1963,Konstantinov:1964,Cowsik:1964,Hara:1980,Hara:1980mz})
that the resonant annihilation of extremely energetic cosmic neutrinos
(EEC$\nu$)---with $E > 10^{20} \ {\rm eV}$---with relic anti-neutrinos
(and vice versa) into $Z$-bosons appears
to be a unique process having sensitivity to the relic neutrinos.
On resonance,
\begin{equation}
E^{\rm res}_\nu=\frac{m_Z^2}{2m_\nu }\simeq 4\times 10^{21}\,
\left( \frac{\rm eV}{m_\nu }\right)
 \ {\rm eV},
\end{equation}
the associated cross section is enhanced by several orders of magnitude,
\begin{equation}
\langle\sigma_{\rm ann}\rangle =\int ds \ \sigma_{\nu\bar\nu}^Z(s)/m_Z^2\simeq
2\pi\sqrt{2}\,G_F\simeq 4\times 10^{-32}\ {\rm cm}^2
,
\end{equation}
leading to a ``short'' mean free path $\ell_{\nu}=(\bar n_{\nu}\,
\langle \sigma_{\rm ann}
\rangle)^{-1}\simeq 1.4\times 10^5 \ {\rm Mpc}$ which is {\it only}
about $48\,h$ times the Hubble distance.
Neglecting cosmic evolution effects,
this corresponds to an annihilation probability for EEC$\nu$
from cosmological distances
on the relic neutrinos of $2\,h^{-1}\%$.

The signatures of annihilation are (i) absorption
dips~\cite{Weiler:1982qy,Weiler:1983xx,bib:absorption}
(see also \cite{Gondolo:1991rn,Roulet:1993pz,Yoshida:1997ie})
in the EEC$\nu$  spectrum at the resonant energies, and (ii) emission
features~\cite{Fargion:1999ft,Weiler:1999sh,Yoshida:1998it,Fodor:2001qy,Fodor:2002hy} ($Z$-bursts)
as protons and photons with energies spanning a decade or more above
the Greisen--Zatsepin--Kuzmin (GZK) cutoff at
$E_{\rm GZK}\simeq 4\times 10^{19}$~eV~\cite{Greisen:1966jv,Zatsepin:1966jv}.
This is the energy beyond which the CMB is absorbing to nucleons due to
resonant photopion production.\footnote{The association of $Z$-bursts with
the mysterious
cosmic rays observed above $E_{\rm GZK}$ is a controversial
possibility~\cite{Fargion:1999ft,Weiler:1999sh,Yoshida:1998it,Fodor:2001qy,Fodor:2002hy,
Kalashev:2001sh,Gorbunov:2002nb,Semikoz:2003wv,Gelmini:2004zb}.}

The possibility to confirm the existence of relic neutrinos within
the next decade from a measurement
of the aforementioned dips in the EEC$\nu$
flux was recently investigated in \cite{bib:absorption}.
Presently planned neutrino detectors (Pierre Auger Observatory~\cite{Auger},
IceCube~\cite{IceCube}, ANITA~\cite{ANITA}, EUSO~\cite{EUSO},
OWL~\cite{OWL}, and SalSA~\cite{Gorham:2001wr})
operating in the energy regime above $10^{21} \ {\rm eV}$ appear to
be sensitive enough
to lead us, within the next decade, into an era of relic neutrino absorption
spectroscopy, provided that the EEC$\nu$ flux
at the resonant energies is close to current observational
bounds and the neutrino mass
is
$>0.1 \ {\rm eV}$. In this case, the associated $Z$-bursts
must also be seen as post-GZK events at the planned cosmic ray detectors
(Auger, EUSO, and OWL).

What are the implications of relic neutrino clustering
for absorption and emission spectroscopy?
Firstly, absorption spectroscopy is predominantly sensitive to the relic
neutrino background at early times, with
the depths of the absorption dips determined largely by the higher number
densities at $z\gg 1$.
Since neutrinos do not cluster significantly
until $z< 2$, clustering at recent times can only show up as
secondary dips with such minimal widths in energy~\cite{Reiter:unpubl}
that they do not seem likely to be resolved by planned observatories.

\begin{figure}[t]
\centering
\includegraphics[width=11.5cm]{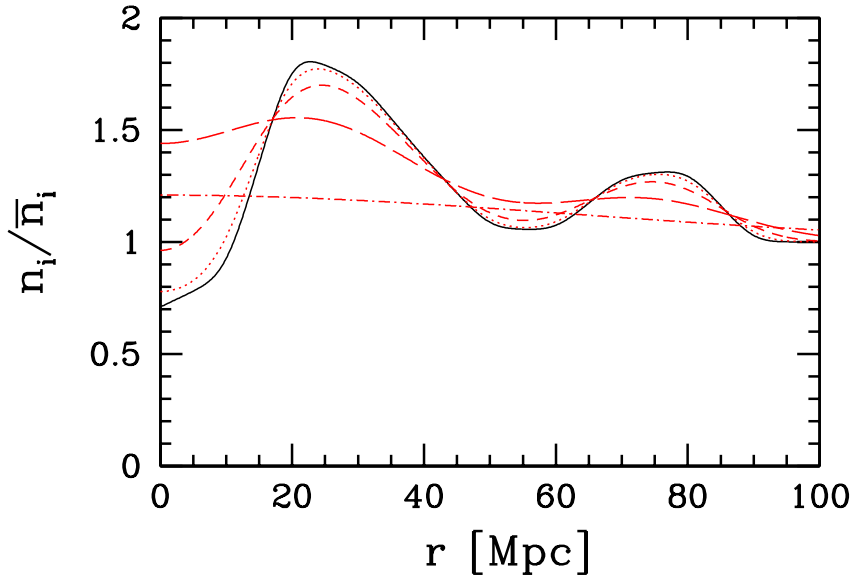}
%
%
\caption{\label{fig:local}``Large scale'' overdensities ($i=\nu,{\rm CDM}$) in the
local universe, with the Milky Way at $r=0$.
The black (solid) line corresponds to the local CDM distribution
inferred from peculiar velocity measurements
\cite{daCosta:1996nt} (see also \cite{Dekel:1998we})
smeared over the surface of a sphere with radius $r$ \cite{Fodor:2002hy}.
The dotted line is the neutrino overdensity for
$m_{\nu} = 0.6 \ {\rm eV}$, short dash $0.3 \ {\rm  eV}$, long dash
$0.15 \ {\rm eV}$, and dot-dash $0.04 \ {\rm eV}$.}
\end{figure}

On the other hand, emission spectroscopy is directly sensitive to the
relic neutrino content
of the local universe ($z< 0.01 \Leftrightarrow
r_{\rm GZK} < 50 \ {\rm Mpc}$).
However, since the neutrino density contrasts approximately track those of
the underlying CDM above the neutrino
free-streaming scale $k_{\rm fs}^{-1}$,
it is clear that there cannot be a substantial neutrino overdensity over the
whole GZK volume ($\sim r_{\rm GZK}^3$).  Indeed, given
the local CDM distribution inferred from peculiar velocity measurements
(smeared over $\sim 5 \ {\rm Mpc}$), we estimate the corresponding
neutrino overdensity to be $< 2$ (Figure~\ref{fig:local}).
Hence the overall emission
rate cannot be significantly enhanced by gravitational clustering.

Another possibility is to  exploit the fact that there are several
galaxy clusters ($\! >  10^{14} M_{\odot}$) within the GZK zone with significant
neutrino clustering.
One could then search for
directional dependences in the emission events as a signature of
EEC$\nu$--relic $\nu$ annihilation.
For example, AGASA has an angular resolution of $\sim 2^{\circ}$
\cite{bib:agasa}. This is already sufficient to resolve the internal
structures of, say, the Virgo cluster
(distance $\sim 15 \ {\rm Mpc}$,
$M_{\rm vir} \sim 8 \times 10^{14} M_{\odot}$)
which spans some $10^{\circ}$ across the sky.
From Figure~\ref{fig:overdensities},
the average neutrino overdensity along the line of
sight towards and up to Virgo is estimated to be
$\sim 45$ and $\sim 5$ for $m_{\nu} = 0.6 \ {\rm eV}$ and
$0.15 \ {\rm eV}$ respectively, given an angular resolution of
$\sim 2^{\circ}$.  The corresponding increases in the number of events
coming from the direction of the Virgo cluster relative to the unclustered
case, assuming an isotropic distribution of EEC$\nu$ sources,
 are given roughly by the same numbers, since protons originating
from $\sim 15 \ {\rm Mpc}$ away arrive at Earth approximately unattenuated.
The numbers improve to $\sim 55$ and $\sim 8$ respectively with a finer
$\sim 1^{\circ}$ angular resolution.

\section{\label{conclusion} Conclusion}

We have conducted a systematic and exhaustive study of the gravitational
clustering of big bang relic neutrinos onto existing CDM and
baryonic structures
within the flat $\Lambda$CDM model.
Our main computational tools are (i) a restricted, $N$-one-body method,
in which we neglect the gravitational interaction
between the neutrinos and treat them as test particles moving in an
external potential generated by the CDM/baryonic structures,
and (ii) a semi-analytical, linear technique,
which requires additional assumptions about the neutrino phase space
distribution.  In both cases, the CDM/baryonic gravitational
potentials are calculated from parametric halo density profiles from
high resolution $N$-body studies  and/or from
realistic mass distributions reconstructed from observational data.

Using these two techniques, we track
the relic neutrinos' accretion onto CDM halos ranging from the galaxy to the
galaxy cluster variety ($M_{\rm vir} \sim 10^{12} \to 10^{15} M_{\odot}$),
and determine the neutrino number densities
on scales $\sim 1 \to 1000 \ {\rm kpc}$ for a range of neutrino masses.
We find that the linear method systematically
underestimates the neutrino overdensities over the whole range of
halo and neutrino masses considered.
Reconciliation with $N$-one-body simulations can only be achieved if
we impose a smoothing scale of $> 1 \ {\rm Mpc}$, or if
the overdensity is no more than three or four.
We therefore conclude that the linear theory does not generally
constitute a faithful approximation to the Vlasov equation
in the study of neutrino clustering on  galactic and  sub-galactic scales
($\! < 50 \ {\rm kpc}$).  However, it may still be useful for finding the
minimum effects of neutrino clustering in other contexts not considered in this
work (e.g., the nonlinear matter power spectrum \cite{bib:kevtalk}).

Next we estimate the neutrino phase space distribution in the Milk Way,
especially in our local neighbourhood at Earth $r_{\oplus}$,
 taking also into
account contributions to the total gravitational potential
from the galactic bulge and disk.
We find a maximum overdensity of $\sim 20$ per neutrino flavour
in our immediate vicinity, provided that the neutrino mass is
at its current upper limit of $0.6 \ {\rm eV}$.
For neutrino masses less than
$0.15 \ {\rm eV}$, the expected overdensity from gravitational clustering
is less than two.
The associated coarse-grained momentum spectra show varying degrees of
deviation from the relativistic Fermi--Dirac function,
but share a common feature that they are semi-degenerate, with phase space
density $\bar{f} \sim 1/2$,  up to the
momentum state corresponding to the escape velocity from the
Milky Way at $r_{\oplus}$.  This means that the neutrino number densities
we have calculated here for $r_{\oplus}$ are already the
{\it highest possible}, given the neutrino mass, without violating
phase space constraints.
In order to attain even higher densities, one must now appeal to
non-standard theories (e.g., \cite{Stephenson:1996qj}).

In terms of scattering-based detection possibilities, this meager enhancement
in the neutrino number density in the Milky Way from gravitational clustering  
means that relic neutrinos are still far from being detected in fully
earthbound laboratory experiments.  For flux detection methods 
based on coherent elastic scattering of relic neutrinos off
target matter in a terrestrial detector, a
positive detection could be thirty to forty years away,
provided that light neutrinos are Dirac particles.  For light
Majorana neutrinos,
another $\sim 10^3$ times more sensitivity would be required in the detector
for a positive signal.
Target detection methods using accelerator beams
seem equally hopeless, unless the accelerator is the size of the Earth
and operates at an energy of $\sim 10^7 \ {\rm TeV}$.

Meanwhile, target detection using extremely energetic cosmic neutrinos 
(EEC$\nu$, $\! > 10^{21} \ {\rm eV}$) remains the only viable means to
confirm the existence of big bang relic neutrinos
within the next decade or so.  Resonant
annihilation of EEC$\nu$ on relic neutrinos can be revealed as
absorption dips in the EEC$\nu$ flux (e.g., \cite{bib:absorption}), or
as emission features in the $Z$-decay products.  However, since absorption 
spectroscopy is largely insensitive to late time ($z < 2$)
relic neutrino clustering,
our findings here have little impact on the conclusions of
\cite{bib:absorption}.  On the other hand, emission spectroscopy is
sensitive to the relic neutrino content of the local GZK zone,
$V_{\rm GZK} \sim 50^3 \ {\rm Mpc}^3$.  
While we find no significant large scale clustering
within $V_{\rm GZK}$ and therefore no significant
enhancement in the overall emission rates, it is still conceivable to
exploit the considerable neutrino overdensities in nearby galaxy clusters, 
and search for directional dependences in the  post-GZK emission events.
For the Virgo cluster, for example, we estimate the event rate from
the central $1^\circ$ region to be
$\sim 55$ and $\sim 8$ times the unclustered rate for neutrino mass
$m_{\nu} = 0.6 \ {\rm eV}$ and $0.15 \ {\rm eV}$ respectively, assuming
an isotropic distribution of EEC$\nu$ sources.
Planned observatories such as the Pierre Auger
Observatory \cite{Auger},  EUSO \cite{EUSO} and OWL \cite{OWL} 
will have sufficient angular resolution to, in principle, see this
enhancement. However, considering the rapidly improving constraints
on both the EEC$\nu$ flux and neutrino masses, it remains to be
seen if the enhancement can indeed be observed 
with enough statistical significance \cite{bib:virgo}.

%
%

%
%



\printindex
\end{document}